\documentclass[]{spie}  %>>> use for US letter paper
\usepackage{pdfpages}
\usepackage{amsmath,amsfonts,amssymb}
\usepackage{mathtools}
\usepackage{graphicx}
\usepackage[colorlinks=true, allcolors=blue]{hyperref}

\title{Segmentation of kidney stones in endoscopic video feeds}

\author[a]{Zachary A Stoebner}
\author[a]{Daiwei Lu}
\author[a]{Seok Hee Hong}
\author[b]{Nicholas L Kavoussi}
\author[a]{Ipek Oguz}
\affil[a]{Vanderbilt University, Nashville, TN, USA}
\affil[b]{Vanderbilt University Medical Center, Nashville, TN, USA}

\begin{document}
\maketitle

\begin{abstract}
Image segmentation has been increasingly applied in medical settings as recent developments have skyrocketed the potential applications of deep learning. Urology, specifically, is one field of medicine that is primed for the adoption of a real-time image segmentation system with the long-term aim of automating endoscopic stone treatment. In this project, we explored supervised deep learning models to annotate kidney stones in surgical endoscopic video feeds. In this paper, we describe how we built a dataset from the raw videos and how we developed a pipeline to automate as much of the process as possible. For the segmentation task, we adapted and analyzed three baseline deep learning models -- U-Net, U-Net++, and DenseNet -- to predict annotations on the frames of the endoscopic videos with the highest accuracy above 90\%. To show clinical potential for real-time use, we also confirmed that our best trained model can accurately annotate new videos at 30 frames per second. Our results demonstrate that the proposed method justifies continued development and study of image segmentation to annotate ureteroscopic video feeds.
\end{abstract}


\section{INTRODUCTION}
\label{sec:intro}

\subsection{Background}

In recent years, image segmentation, i.e., the automatic identification of the location and boundary of objects within an image, has proven useful in myriad applications, from medical image analysis to robotic perception to self-driving cars. With increased applicability, extensive research is underway to develop better image segmentation techniques, particularly those that use deep learning \cite{Minaee2021}. 

Even more recently, the general deep learning architecture that has proved especially useful to the task of image segmentation is the encoder-decoder network \cite{Zaitoun2015}. These networks learn to encode the image into a low-dimensional latent vector and then decode the latent vector to recreate the image with different attributes or, in the case of segmentation, recreate a mask of the image encoded with the class of each pixel. Example architectures that changed the landscape of image segmentation include SegNet \cite{Badrinarayanan2017} for scene understanding, U-Net \cite{Ronneberger2015a} for biomedical image segmentation, and DeepLabv3 \cite{Chen2017} for multi-scale segmentation. 

SegNet contributed max unpooling layers to mirror VGG-16’s encoder-like architecture \cite{Simonyan2015} to generate segmentation masks. DeepLabv3 pioneered atrous convolution to capture multi-scale context in images. Of particular interest, U-Net demonstrated surprisingly robust segmentation of biomedical images and many recent medical image segmentation models are variants of this architecture. U-Net has a typical deep encoder-decoder structure but complements a simple feed-forward network with skip connections from each layer of the encoder to each corresponding layer of the decoder, mitigating the vanishing gradient problem. Variants of U-Net, such as U-Net++ \cite{Zhou2020}, propose improvements to extend the characteristic U-Net architecture to compose deeper, yet more robust models. 

\subsection{Motivation}

In the medical field, many surgical specialties have yet to benefit from deep learning technology mainly due to the sparsity of available datasets for training, leading to slow adoption\cite{pmid30514693}. However, as these technologies develop, many specialties are beginning to investigate applications of deep learning tools to augment clinical practice. Due to the prevalence of minimally invasive surgical techniques in urology, the field is well-positioned for capitalizing on image segmentation to automatically detect kidney stones in video feeds \cite{Nithya2020,Viswanath2014}. Specifically, image segmentation can be used as a surgical aid and the ability to process real-time feeds can motivate future growth towards an automated image-guided surgery system.

In this project, we explore the automated annotation of kidney stones from endoscopic video feeds with supervised deep learning-based image segmentation methods. Our method focuses on semantic segmentation (i.e., the pixel-wise detection of class belonging), instead of instance segmentation (i.e., the pixel-wise identification of instance belonging). We aim to establish the feasibility of real-time annotation of kidney stone video feeds in this project, with the long-term goal of building an image-guided surgical system for clinicians and eventually robotic agents. To meet these goals, we built a novel dataset from surgical endoscopic video feeds and investigated models and techniques to generate accurate segmentations. We explored three baseline models for this purpose: U-Net, U-Net++, and DenseNet \cite{Jegou2017,Huang2017}. 

\section{METHODS}
\label{sec:methods}

\subsection{Dataset and Preprocessing}

We received raw ureteroscopy videos of both digital and fiberoptic formats; we opted to only focus on the digital videos in this initial study due to their higher quality. After quality assuring the raw inputs by visual inspection, we extracted 20 frames per second [FPS] from the videos. We cropped each image by first converting to grayscale and using Otsu thresholding \cite{1979:ots} to separate the background from the foreground, then finding the contours using OpenCV and selecting the closed contour with the greatest area to identify a bounding box for the image. Compared to manual cropping and non-parallelized cropping algorithms, this method is extremely efficient, capable of processing videos at their playback speed. Out of a total of 29 videos, two were removed due to text overlaid on the scope's video after capture by the surgeon, which thwarted clean cropping of these videos.

\begin{figure}[h]
\centering
\includegraphics[scale=0.7]{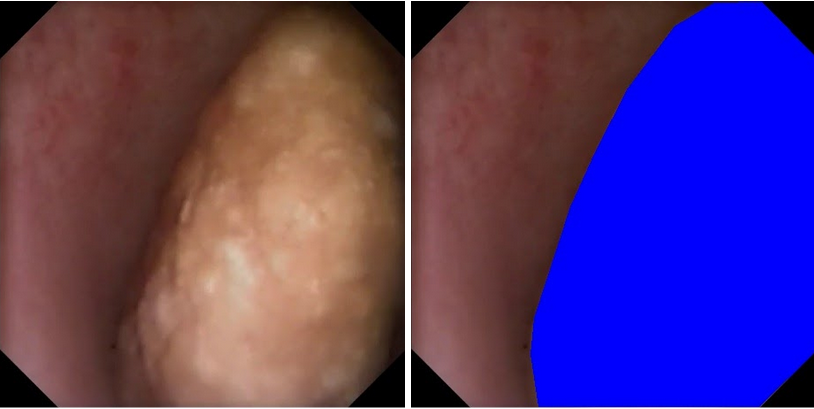}
\caption{Example of a cropped image and its manual annotation. The images were annotated using MakeSense.ai and the vertices of the polygon drawings were saved in JSON files. After parsing the JSON files, the polygons were then converted into binary images and displayed as an overlay on original images using OpenCV to visualize the annotation. 
} 
\label{fig1}
\end{figure}

After cropping, the frames were manually annotated for kidney stones to generate the ground truths for the examples in the dataset, both for training and evaluation purposes. To annotate the cropped images, we employed MakeSense.ai, a web application that supports polyline annotations and saves the data in JSON format. To pair example images with their annotated ground truths in the dataset, we used OpenCV to convert the polygons into binary images. These pairs of images were used for supervised training of the models. Figure \ref{fig1} displays an example of a cropped image and its manual annotation.

% \begin{table}[h]
%     \centering
%     \includegraphics[width=0.5\textwidth]{splits.jpg}
%     \caption{Summary of the dataset split between training and testing sets. Out of 27 videos, 22 videos were utilized in training and 676 frames from these videos were annotated. At the start of training, 10\% of the training frames were randomly split for the validation set; hence, approximately 67 frames were used for validation.}
%     \label{tab1}
% \end{table}

\subsection{Architectures and Training}
We employed three architectures, U-Net, U-Net++, and DenseNet, to establish baselines of current state-of-the-art image segmentation models for the kidney stone segmentation task. 
The main baseline that we used was a U-Net with a depth of 5 and composed of ResNet34 blocks \cite{He2016}. Similar architecture parameters were applied to our U-Net++ model; however, it is important to note that U-Net++ contains numerous subnetworks that complicate the model but also make it more robust. Hence, U-Net++ has significantly more parameters than a typical U-Net. The third type of model that we looked at was a DenseNet \cite{Jegou2017}, which implements the U-Net architecture but with DenseBlocks  \cite{Huang2017} instead of ResNet blocks. The variant that we focused on was DenseNet67. Each variant had a depth of 5 with different blocks sizes, growth rates, and bottleneck sizes.   

Training was a standard iterative pipeline with forward propagation and loss computation, followed by backpropagation with frequent validation checks. The training and validation sets were determined randomly based on the number of videos, where the number of videos in the testing set is 10-20\% of the total number of videos. Out of 27 videos, 22 videos were thus utilized in training and 676 frames from these videos were annotated. The remaining five videos comprise the test set with a total of 73 annotated frames; the hold-out test set contains many examples of common challenges in the data, i.e., motion blur, debris fragmentation, foreign objects, and saline injection. The model does not update its parameters on the validation data as no loss is computed for the validation set. Rather, it is used in a similar fashion as the test set to generate scores and segmentation masks throughout training to better monitor and understand the model's performance. We also compared the results of vanilla and pretrained U-Net and U-Net++ models to confirm whether pretraining is helpful for this task. To confirm consistency in performance, we have trained multiple copies of the best hyperparameter configuration found for each architecture. In total, we have saved 8 high-performing models for U-Net++, 6 for U-Net, and 3 for DenseNet. On top of that, we have archived the many more models acquired from the grid search on each architecture. 

%Table \ref{tab1} summarizes the size of the dataset and the train / test splits.

\subsection{Statistics}

To robustly identify the best models and their best hyperparameters, we conducted a standard grid search for each model and plotted their training and loss curves. To compare the accuracies of the different architectures throughout training, testing, and searching, we computed the Sorensen-Dice coefficients \cite{s1948a}. At the image level, the Dice score is twice the number of pixels overlapping between the prediction and the ground truth, divided by the total number of segmented pixels across both images. We chose the Dice score as our primary measure of performance because it is well-defined for balanced, binary-classed datasets and many deep learning methods for semantic segmentation also measure the Dice score. 

$$ Dice(P,G) = \frac{2|P\cap G |}{|P| + |G|} \textrm{, where P is the predicted set and G is the target set. }$$

Other statistics commonly applied to image segmentation include the pixelwise accuracy,  intersection over the union (IoU), peak signal-to-noise ratio (PSNR), receiver operating characteristic (ROC), and area under the curve (AUC). These metrics are supplementary and also yield the rates for true positives, true negatives, false positives, and false negatives. For our loss function, we used binary cross-entropy (BCE).

\subsection{Implementation Details}
Beyond the utilization of OpenCV to handle the video and image data, we heavily employed PyTorch and Comet.ml to carry out our model training and analysis. PyTorch comprised the deep learning framework for our models and Comet.ml was used to extract relevant metrics and images throughout the training and testing. Comet.ml then logged all of the information to a web browser workspace, aggregating all of our experiments and collected data and plotting the live curves of the model during training. Data points were collected at each step in the training process where $total\ steps\ =\ \lceil\frac{dataset\ length}{batch\ size}\rceil\ \ast\ epochs$.

\section{RESULTS}
\label{sec:results}

\subsection{Pretraining}

\begin{figure}[h]
\includegraphics[width=\textwidth]{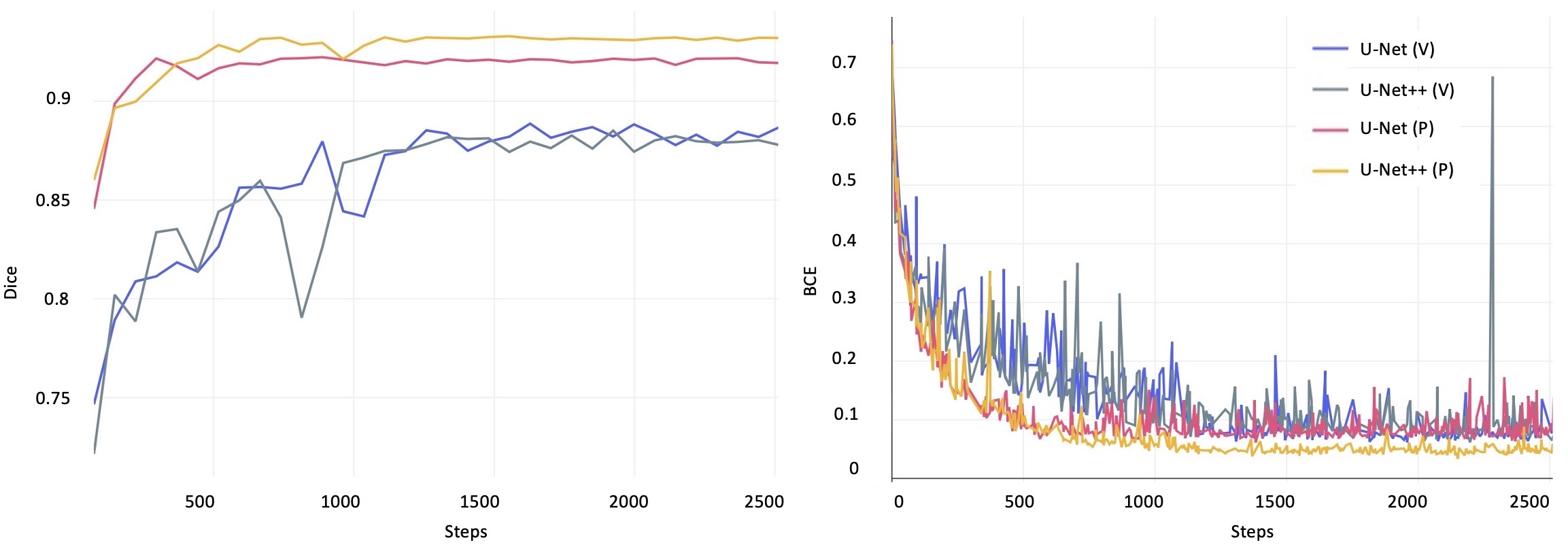}
\caption{A comparison of vanilla (V) and pretrained (P) model training accuracies. Vanilla U-Net and U-Net++ start at a lower accuracy of approximately 0.75 Dice and gradually converge to a Dice score of approximately 0.88. ImageNet-pretrained U-Net and U-Net++ start at approximately 0.85 Dice, the convergence of the vanilla versions, and converge to about 0.91 and 0.92 Dice, respectively.   
} \label{fig2}
\end{figure}

As expected, we confirmed that pretraining our U-Net models on ImageNet sped training up by approximately 80\%. In essence, the pretrained models have a head start and yield higher accuracies early in training and converge to their maximal accuracies shortly thereafter. The training curve comparison between a vanilla U-Net and ImageNet-pretrained U-Net and U-Net++ are displayed in Figure \ref{fig2}. Our DenseNet models were not pretrained; regardless, DenseNet executed much more slowly on a GPU than the other two architectures, often taking hours to train while the U-Net and U-Net++ took minutes to train.  

\subsection{Training the Baseline Models}

\begin{figure}[h]
\centering
\includegraphics[width=\textwidth]{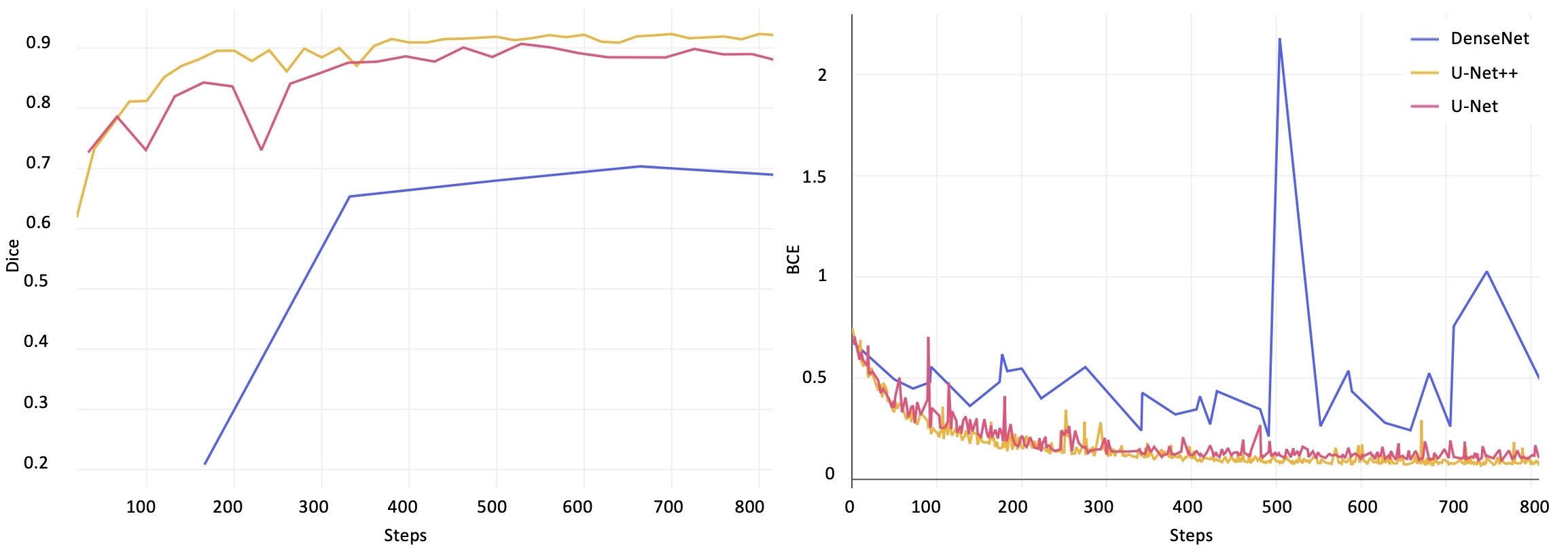}
\caption{A comparison of our best models’ Dice score (left) and BCE loss (right) from training. Note that the y-axis is scaled to the range of values. DenseNet had the highest and most variant BCE loss, yet its outlier values are relatively low compared to those in other binary classification tasks as BCE does not have an upper bound.  
} \label{fig3}
\end{figure}

Quantitative results for our three baseline models are shown in Figure \ref{fig3}, with the set of hyperparameters that led to the best accuracy for each architecture. We observed that the ImageNet-pretrained U-Net++, with a batch size of 8 and learning rate of 5e-5, performed the best on average with the highest Dice scores reaching above 0.91, followed by the ImageNet-pretrained U-Net, with a batch size of 4 and a learning rate of 1e-4, ranging in Dice score between 0.8-0.9, and finally the ImageNet-pretrained DenseNet67, with a batch size of 1 and learning rate of 1e-4, which capped out at a Dice score of approximately 0.7.  Performance often converged within a ± 0.1 Dice of the mean score of a certain architecture. Additionally, we observed that our models were highly sensitive to the learning rate and normalization. Low learning rates and no normalization achieved the best performance. Table \ref{tab2} summarizes the maximum validation statistics, in parentheses, for each baseline model. 

With these findings, we determined that the ImageNet-pretrained U-Net++ is the best model for the kidney stone segmentation task, so further experimentation will continue in this direction. 

\subsection{Real-time Video Feed and Qualitative Results}

\begin{figure}[h]
\centering
\includegraphics[width=\textwidth]{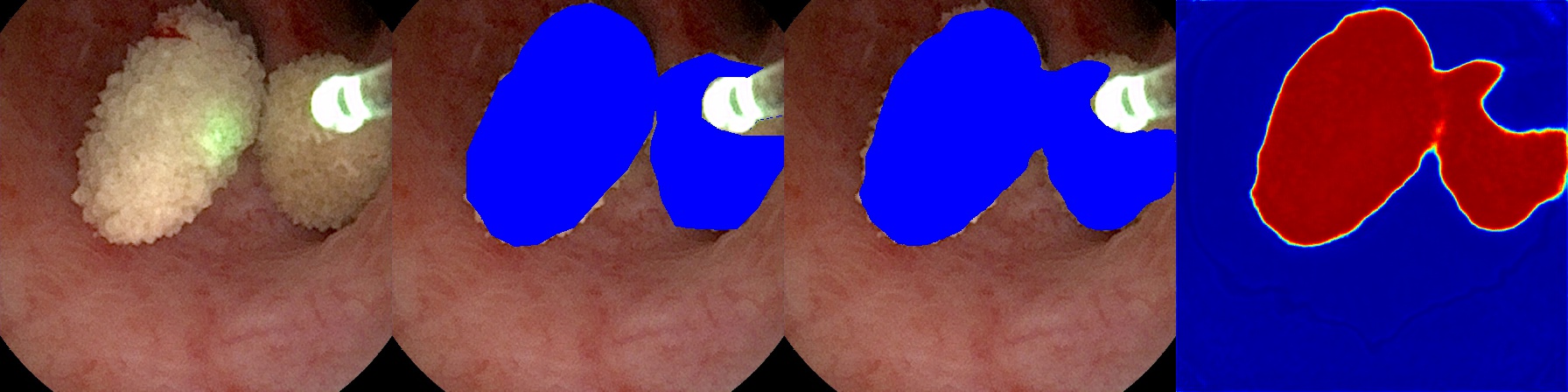}
\caption{Sample frame from the side-by-side video reconstruction of the input, ground truth, automated prediction with U-Net++, and heat map (left to right). The heat map is the raw probability output per pixel whereas the predicted segmentation is the pixels with probabilities $\geq$ 0.5. The model was able to compute this output at 30 FPS. 
} 
\label{fig4}
\end{figure}

We developed a script for processing video feeds and making predictions using our trained models. The script provided sequential frames to the model and ran slightly faster than the duration of each video being processed, which is better than 30 FPS. Videos were separated and reconstructed using OpenCV’s video processing pipeline. Together with the model input and the predicted annotation, we reconstructed the video, containing a model input, its predicted annotation, and the corresponding probability map for side-by-side comparison, from the frames sequentially passed through the model. For testing, ground truth images were also available and included in the comparison image. Figure \ref{fig4} shows an example of a frame used for side-by-side comparison with its ground truth, prediction, and probability shown as a heat map overlay. The GPU hardware used was an NVIDIA RTX 2080 Ti. 

\subsection{Hold-Out Performance and Generalizability}

To evaluate the performance of our model on challenging unseen data, we held out a test set comprised of data with many challenging examples that are common in realistic scenarios, including motion blur, debris fragmentation, foreign objects, and saline injection. Table \ref{tab2} summarizes the test performances of each baseline model, with the highest validation performances in parentheses. 

\begin{figure}[h]
\centering
\includegraphics[width=\textwidth]{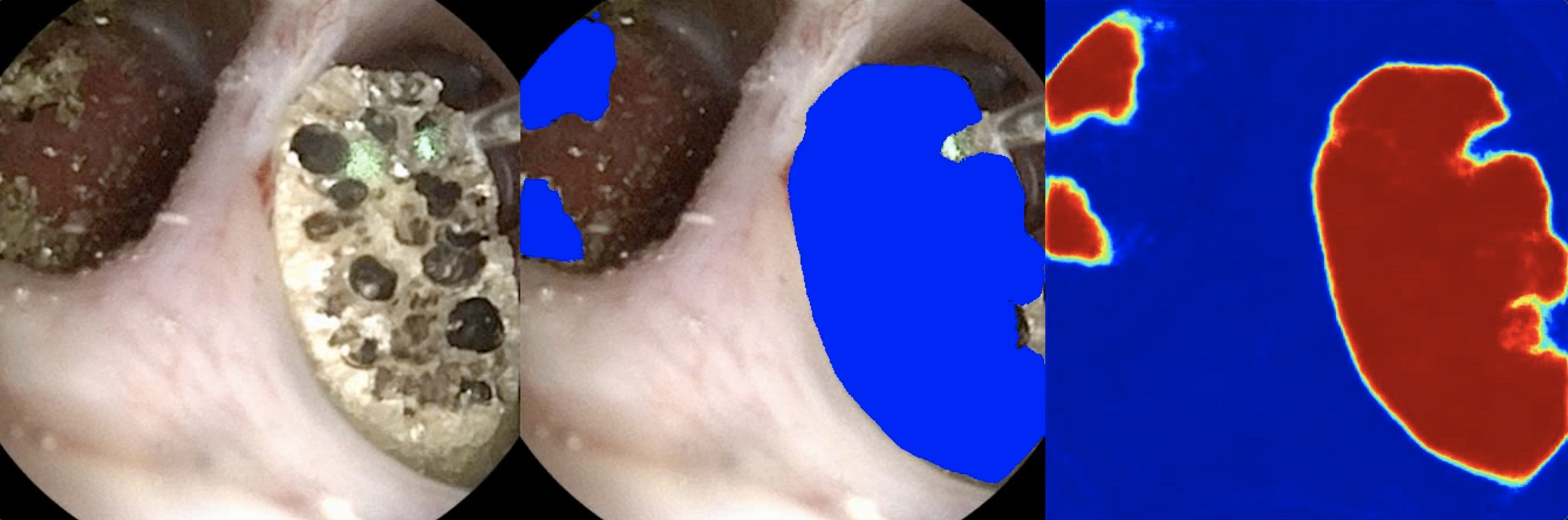}
\caption{Sample output frame for a representative unlabeled input video  from our dataset with our best U-Net++ model. This video has no ground truth annotation at the time of publication. Debris collects in the left duct and is still segmented by the model.
} 
\label{fig5}
\end{figure}

Additionally, we have unlabeled videos in our dataset as mock up examples of realistic videos that a deployed model would see. Shown in Figure \ref{fig5} for a representative unlabeled video in our dataset, we created segmentation predictions for these videos using our video processing script described above as a proof-of-concept for the visual overlay that we intend to deploy in operating rooms where our model will receive real-time and ``in-step'' video frames from the endoscopic hardware. 

\begin{table}[h]
    \centering
    \includegraphics[width=\textwidth]{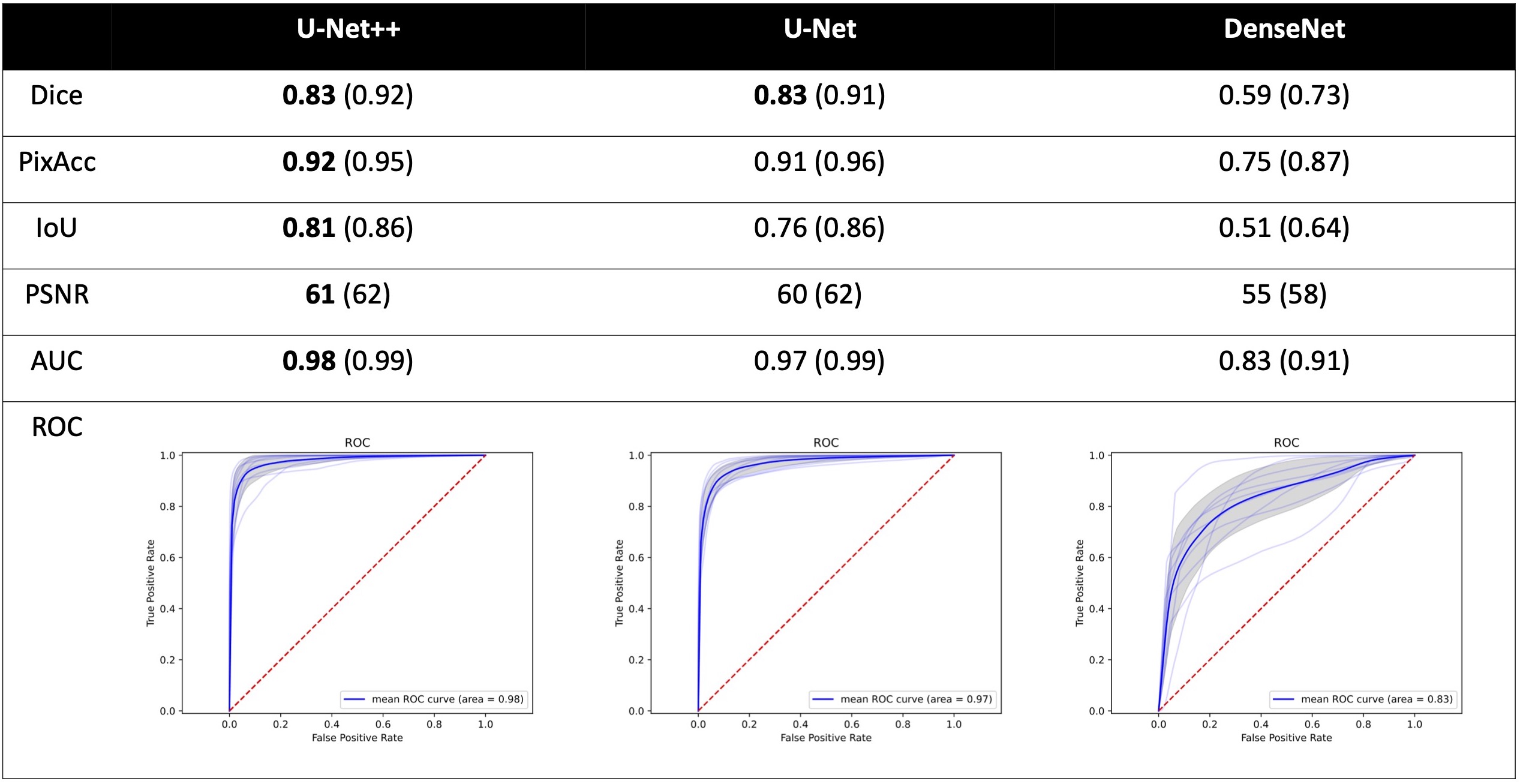}
    \caption{Summary of the statistics gathered for each model. Non-parenthetical values are the average scores from all frames in the test set. The reported value in parentheses is the maximum value recorded from each baseline's validation during training. The highest performances between models for each metric are denoted in bold. U-Net++ claimed the highest scores in each metric on the test set. U-Net had the same test Dice score as U-Net++ but lower scores in all other metrics. DenseNet had relatively poor performance for all metrics. Only the ROC curves from test set performances are included.}
    \label{tab2}
\end{table}

\section{Discussion}
\label{sec:disc}

\subsection{Interpretation of Results}

With relatively few training examples, our U-Net++ and U-Net models achieve an accuracy greater than 90\%. At this performance level, our models can be interpreted by surgeons and used to assist them visually to identify kidney stones in the video feed. With a larger dataset and additional improvements to the U-Net and U-Net++ architectures, we expect that the model will perform better on a wider range of scenarios that occur in realistic endoscopic surgeries. 

The sensitivity of the models to the learning rate and the slightly variable performances of the same models in random restarts suggest that the cost landscape for this task may require multiple  restart attempts for the same model to achieve the best optimization. Given the similarity in performance between U-Net and U-Net++, U-Net may achieve a higher Dice score on some restarts, or with a larger dataset. Although the output videos are real-time, they are not ``in-step'' with the input video, which is passed through the model first and each frame is added to the OpenCV video object to generate the video reconstruction at the end. 

Although the training set still had many examples of motion blur, debris fragmentation, foreign objects, and saline injection, the test set videos had empirically more examples of these kinds. We expect that these challenging frames are particularly common among novice surgeons who could greatly benefit from a tool that visually assists identification of stones. Our high-performing models still score $>$0.8 Dice on this data and, in video reconstructions of unlabeled video. 

 With its current performance, our high-performing model could be used to automatically annotate new kidney stone videos to contribute to dataset expansion, alongside rigorous quality assurance. With improved performance, a visual robotic system employing our model could ideally become an end-to-end automatic solution for urological endoscopic surgeries in the long term. 

\subsection{Clinical Relevance}
Our innovative approach to training a supervised model for tracking kidney stones and integrating this information in the surgical display during endoscopic stone surgery could enhance a surgeon’s ability to diagnose and treat kidneys stones. Due to the limited field of view, visibility during endoscopic stone surgery can be impacted by blood and debris which decreases stone free rates leading to recurrence events \cite{Iremashvili2019}.  Our system could potentially mitigate these visibility issues and improve stone treatment by leveraging these computer vision techniques. Similar applications of deep learning algorithms have shown potential in augmenting surgical technique and safety in robotic and laparoscopic surgeries \cite{WARD20211253}. Furthermore, our system is potentially generalizable since future researchers developing automated tracking and video segmentation systems could use our basic approach for other endoscopic surgeries.

\subsection{Future Improvements}
To deploy our high-performing model for surgical use in operating rooms, we will extend our video processing script to receive ``in-step'' frame-by-frame video input from a video capture card connected via DVI/HDMI to the endoscopic hardware. Then, we will output the side-by-side frames, as in Figure \ref{fig5}, to an adjacent monitor to assist physicians in surgery. Developing such a system will allow us to investigate further goals, such as monocular depth prediction which might prove critical in automation \cite{fu2018deep}.

We will also investigate the application of temporal models for our system. Incorporating information from previous frames might allow for more consistent prediction between subsequent frames. In addition, such models account for memory of data from previous frames without the overhead of additional input dimensions suffered by our current fully convolutional models. For this task, we will incorporate self-attention modules into our U-Net and U-Net++ architectures, which has been shown to increase performance in video segmentation \cite{ji2021progressively}. 

In practice, the problem domain also requires the segmentation of kidney stones after they have been surgically broken down into smaller pieces. The current dataset has been developed to support the segmentation of only whole kidney stones. Further development will include expansion of another section of the dataset where, as a surgeon breaks stones apart, the debris fragments will still be labeled by our model. In Figure \ref{fig5}, our model already segments clumps of debris; however, our future goal is finer granularity via an instance segmentation method \cite{zhou2019cia}. 

Additionally, we plan to incorporate multi-class segmentation to also identify, for example, healthy vs.\ unhealthy tissue. Since the task performs well on stone segmentation, we hypothesize that we can utilize the same underlying architectures to adapt to multi-class segmentation and that the model will perform similarly well, with relatively few manually annotated examples \cite{kayalibay2017cnn}.

\section{Conclusion}
\label{sec:conc}

In this paper, we present an exploratory analysis of automated kidney stone segmentation from ureteroscopic videos using supervised learning models. Our deep learning models achieve promising performance in this novel application domain, which further credits the utility of deep learning in image and video segmentation. With pretraining on ImageNet, we found that U-Net++ is the best-performing model for the task, followed closely by U-Net, while DenseNet performed the worst. Our results indicate the potential of establishing a dataset and training deployable models that could visually assist surgeons in real operating rooms.  Our future research directions will be towards an automated visual control system for, ultimately, fully robotic surgery. Such a system could be widely adopted in surgical practice and potentially improve patient outcomes after surgical endoscopy.

\bibliography{stoneannotationlib} % bibliography data in report.bib
\bibliographystyle{spiebib} % makes bibtex use spiebib.bst

\end{document}